\documentclass[12pt,twocolumn]{article}
\usepackage[a4paper,left=20mm,right=20mm,top=25mm,bottom=25mm,includeheadfoot]{geometry}

\usepackage{mathpazo}
\usepackage{graphicx}
\usepackage{amsmath,textcomp}
\usepackage{makeidx}
\makeindex

\usepackage{sectsty}
	\allsectionsfont{\sffamily\raggedright}
	\sectionfont{\sffamily\large\raggedright}
	\subsectionfont{\sffamily\normalsize\raggedright}
\usepackage{ftnright}

\usepackage{widetext}
\usepackage{flushend}
\usepackage{cuted}
\usepackage{color}
\usepackage{graphicx}
\usepackage{hyperref}
\usepackage{pstricks,amssymb}

\usepackage{fancyhdr}
\begin{document}
\title{\vspace{-2em}\bfseries\sffamily Why no radiation occurs in the case of a uniformly accelerated charge}
\author{\normalsize Ashok K. Singal\\[2ex]
Astronomy and Astrophysics Division, Physical Research Laboratory\\
Navrangpura, Ahmedabad 380 009, India.\\
{\tt ashokkumar.singal@gmail.com}
}
\date{\itshape Submitted on 23-09-2018}
\maketitle

\thispagestyle{fancy}

\begin{abstract}
{\sffamily

We show that in the case of a uniformly accelerated charge, in its instantaneous rest frame, there is only a radial electric field as the acceleration fields strangely get cancelled {\em at all distances} by a transverse term of the velocity fields. Consequently, no electromagnetic radiation will be detected by any observer from a uniformly accelerated charge, even in the far-off zone. This is in contradiction with  Larmor's formula, according to which a uniformly accelerated charge would  radiate power at a constant rate, which is proportional to the square of the  acceleration. On the other hand, the absence of radiation from such a charge is in concurrence with the strong principle of equivalence, where a uniformly accelerated charge is equivalent to a charge permanently stationary in a gravitational field, and such a completely time-static system could not be radiating power at all.
}\\ 
\hrule
\end{abstract}
\section{Introduction}
According to Larmor's formula, an accelerated charge radiates electromagnetic power at a rate proportional to the square of its acceleration \cite{la97,1,2,25}
\begin{equation}
\label{eq:21a}
{\cal P}=\frac{2e^2\dot{\bf v}^{2}}{3c^3} \:.
\end{equation}
However, the net rate of momentum loss to radiation by such a charge is nil 
\begin{equation}
\label{eq:21a1}
\dot{\bf p}=0\:,
\end{equation}
because of the azimuthal symmetry ($\propto\sin^2\phi$) of the radiation pattern, at least in the case of a non-relativistic motion \cite{1,2,25}, and consequently the formula leads to a violation of the energy-momentum conservation law \cite{68d}. 
Moreover, from the strong principle of equivalence, a uniformly accelerated charge is 
equivalent to  a charge permanently stationary in a gravitational field \cite{4} and in such a completely time-static system, there cannot be radiation of power at any instant, let alone a continuous radiation for an indefinite time interval.

From a careful scrutiny of the electromagnetic fields of a uniformly accelerated charge, we shall show that there is no electromagnetic radiation anywhere, not even in the far-off regions. As we will explicitly demonstrate, this happens because the acceleration fields get cancelled neatly by a transverse component of velocity fields, at all distances. As a result, there is no Poynting flux with a term proportional to the square of acceleration,  usually called the radiated power, implying thereby that no electromagnetic radiation takes place from a uniformly accelerated charge.  
\section{Electromagnetic fields of a uniformly accelerated charge -- no evidence of radiation anywhere}
A uniformly accelerated motion usually implies a motion with a constant proper acceleration, say, ${\bf g}$. 
We may assume it to be a one-dimensional motion, as we can always transform to another inertial frame so as to make the component of the velocity vector in a direction perpendicular to the acceleration vector zero. 

We shall now explicitly demonstrate that in the instantaneous rest frame of a charge with a constant proper acceleration, there is a complete cancellation of acceleration fields by a transverse term in the time-retarded velocity fields {\em at all distances}. 

Electromagnetic fields of a charge $e$, moving in one dimension with a proper acceleration ${\bf g}\:(= \gamma^{3} \dot{\bf v})$, can be written for any given time $t$ as \cite{1,2,25,28}
\begin{eqnarray}
\nonumber
{\bf E}&=&\left[\frac{e({\bf n}-{\bf v}/c)}
{\gamma ^{2}r^{2}(1-{\bf n}\cdot{\bf v}/c)^{3}}\right.\\ 
\nonumber
&&\left. +\frac{e\:{\bf n}\times({\bf n}\times
{\bf g})}{\gamma ^{3}rc^2\:(1-{\bf n}\cdot
{\bf v}/c)^{3}}\right]_{t'}\\
\label{eq:1aa}
{\bf B}&=&{\bf n} \times {\bf E}\:,
\end{eqnarray}
where the subscript {\em t'} indicates that quantities within the square bracket are to be evaluated at the corresponding retarded time $t'=t-r/c$.

Now, for a one-dimensional motion with a constant proper acceleration ${\bf g}$, the velocity ${{\bf v}}$ at the retarded time $t-r/c$ is obtained from its {\em present} value ${{\bf v}}_{\rm o}$ at time $t$ (with $\gamma$, $\gamma_0$ being the corresponding Lorentz factors) as 
\begin{equation}
\label{eq:11c2}
\gamma {\bf v}=\gamma_0 {\bf v}_0- {\bf g}r/c\:.
\end{equation}
Therefore, in the instantaneous rest-frame (${{\bf v}}_0=0$), the proper acceleration and the {\em retarded value} of the velocity are related by $\gamma {\bf v}=-{\bf g}r/c$. Basically this happens because for larger $r$, we need to go further back in time to get the time-retarded value of velocity, which, in the case of a uniform acceleration, is directly proportional to the time interval $r/c$.
Substituting for ${\bf g}$ in Eq.~(\ref{eq:1aa}), and after a rearrangement of terms, we get the electric field in the instantaneous rest-frame as
\begin{eqnarray}
\label{eq:1aaa}
{\bf E}=\left[\frac{e({\bf n}-{\bf v}/c-{\bf n}\times\{{\bf n}\times {\bf v}/c\})}{\gamma ^{2}r^{2}(1-{\bf n}\cdot{\bf v}/c)^{3}} \right]_{t'}.
\end{eqnarray}

Using the vector identity ${\bf n}\times({\bf n}\times{\bf v})={\bf n}({\bf n}.{\bf v})-{\bf v}$, we get the expression for the electric field in the instantaneous rest-frame of a uniformly accelerated charge as
\begin{eqnarray}
\label{eq:11a}
{\bf E}=\left[\frac{e{\bf n}}{\gamma ^2 r^2(1-{\bf n}\cdot{\bf v}/c)^2}\right]_{t'},
\end{eqnarray}
where there is only a radial electric field with respect to the charge position at the  retarded time, with the transverse acceleration fields in Eq.~(\ref{eq:1aa}) having got cancelled by transverse component of velocity fields, for all $r$. There is thus neither any magnetic field nor Poynting flux {\em anywhere} and therefore no observer would detect any radiation at whatever distance, in the instantaneous rest frame. 

This in turn is in agreement with the strong principle of equivalence where a charge permanently stationary in a gravitational field, and thereby with no whatsoever temporal variations, and which is equivalent to a charge having a constant proper acceleration \cite{4}, cannot be continually radiating \cite{17,18}.

At any other time, when ${{\bf v}}_{\rm o}\ne0$, in addition to the radial term in Eq.~(\ref{eq:11a}), we also have a transverse term for the electric field
\begin{eqnarray}
\nonumber
{\bf E}&=&\left[\frac{e{\bf n}}{\gamma ^2 r^2(1-{\bf n}\cdot{\bf v}/c)^2}\right.\\
\label{eq:11aaa}
&&\left.+\frac{e{\bf n}\times\{{\bf n}\times \gamma_0{\bf v}_0/c\}}{\gamma ^{3}r^{2}(1-{\bf n}\cdot{\bf v}/c)^{3}} \right]_{t'}.
\end{eqnarray}
Now the transverse terms, proportional to the present velocity $\gamma_0{\bf v}_0$, fall rapidly with distance ($\propto 1/r^2$); the Doppler beaming factor $\delta^3=1/\gamma^3(1-{\bf n}\cdot{\bf v}/c)^3$ merely redistributing the field strength in solid angle without affecting the net Poynting flux at any $r$. From  Poynting vector, ${\bf S}= c({\bf E}\times {\bf B})/{4\pi}$, 
Eq.~(\ref{eq:11aaa}) yields for the Poynting flux through a spherical surface, $\Sigma$ of radius $r$ \cite{18}
\begin{equation}
\label{eq:11aab}
{\cal S}=\int_{\Sigma}{{\rm d}\Sigma}\: ({\bf n} \cdot {\bf S}) =\frac{2e^{2}}{3 c r^{2}}(\gamma_0 v_0)^{2}.
\end{equation}
We see that the Poynting flux falls rapidly with distance (${\cal S} \rightarrow 0$ as $r\rightarrow \infty$). 
Here we find no term independent of $r$ and proportional to $\dot{\bf v}^{2}$, that is usually defined as the radiated power, implying  thereby, no radiation from a uniformly accelerated charge.

To fully comprehend its physical implications, we replace the uniformly accelerated charge at its position at the retarded time $t'$, by a charge moving with a uniform velocity $v_0$ (equal to the ``present  velocity'' of the accelerated charge). The electric field for such a charge can be written as
\begin{eqnarray}
\nonumber
{\bf E}&=&\left[\frac{e({\bf n}-{\bf v}_0/c)}{\gamma_0 ^{2}r^{2}(1-{\bf n}\cdot{\bf v}_0/c)^{3}} \right]_{t'}\\
\nonumber
&=&\left[\frac{e{\bf n}}{\gamma_0 ^{2} r^{2}
(1-{\bf n}\cdot{\bf v}_0/c)^{2}}\right.\\
\label{eq:11aac}
&&\left.+\frac{e{\bf n}\times\{{\bf n}\times \gamma_0{\bf v}_0/c\}}{\gamma_0 ^{3}r^{2}(1-{\bf n}\cdot{\bf v}_0/c)^{3}} \right]_{t'},
\end{eqnarray}
where we have used the vector identity ${\bf v}_0={\bf n}({\bf n}.{\bf v}_0)-{\bf n}\times({\bf n}\times{\bf v}_0)$. 

Now, in this case too we get the same Poynting flux (Eqs.(\ref{eq:11aab})) through a spherical surface around the retarded position of the charge. This is true for any $r$. As for a charge moving with a uniform velocity, a finite Poynting flux certainly does not imply power being radiated away from the charge, it is merely due to the movement of the charge along with its self-fields, with a velocity $v_0$ with respect to the retarded position upon which the spherical surface is centred. Same is the case of a uniformly accelerated charge which has a ``present'' velocity $v_0$ with respect to its retarded position and therefore it does not represent any power being radiated away from the charge. 
\section{What is amiss in Larmor's radiation formula?}
A pertinent question that could arise here is: If a uniformly accelerated charge does not radiate, which contradicts Larmor's formula, does it mean that Larmor's formula is invalid? How could this issue be resolved?

Actually, in the text-book derivation of Larmor's formula (Eqs.(\ref{eq:21a})), Poynting's theorem is improperly applied to equate the radiated power at time $t$ to the rate of loss of the mechanical energy $({\cal E_{\rm me}})$ of the charge at a retarded time $t'=t-r/c$ 
\begin{equation}
\label{eq:23b}
\left[\frac{{\rm d}{\cal E_{\rm me}}}{{\rm d}t}\right]_{t'}
=- \left(\int_{\Sigma}{{\rm d}\Sigma}\:({\bf n} \cdot {\bf S})\right)_{t} \;.
\end{equation}
However, in Poynting's theorem {\em all quantities} are supposed to be calculated for the {\em same instant of time} \cite{1,2,25}, say, $t$ and one cannot directly calculate the rate of loss of the mechanical energy $({\cal E_{\rm me}})$ of the charge at a retarded time $t'=t-r/c$ from the radiated power at time $t$.

That Eq.~(\ref{eq:23b}) could lead to wrong conclusions can be seen by applying it to the case of an accelerated charge that is instantly stationary at $t'$. 
The charge has no velocity at that instant and hence no kinetic energy, therefore the 
left hand side can only be zero, while the right hand side yields a finite result 
for the Poynting flux (proportional to square of the acceleration of the charge, with no dependence on velocity).
Therefore the derivation of Larmor's formula (Eqs.(\ref{eq:21a})) employing  Eqs.(\ref{eq:23b}) may not be a legitimate one and it is this oversight which could mostly be responsible for the confusion in this century-old problem. 

It is important to note that in the case of a periodic motion of period $T$, there is no difference in the radiated power integrated or averaged between $t$ to $t+T$ and  $t'$ to $t'+T$, therefore Eq.(\ref{eq:23b}), and thereby Larmor's formula, does yield a correct average power loss by the charge for a periodic case. 
Further, for a periodic motion, e.g., a harmonically oscillating charge in a radio antenna, it is easily verified that $<\dot{\bf v}^2>=<-\ddot{\bf v}\cdot{\bf v}>$ \cite{68a}, therefore Larmor's formula (Eqs.(\ref{eq:21a}) yields the same time-averaged radiative power as from the formula derived from 
the famous Abraham-Lorentz radiation reaction formula \cite{abr04,abr05,lor04,16,68b}, where one gets instantaneous power loss of the charge (in a non-relativistic motion) as \cite{68a}
\begin{equation}
\label{eq:6.1}
{\cal P} =-\frac{2e^{2}}{3c^{3}}\ddot{\bf v}\cdot{\bf v}\:.
\end{equation}
However, in a non-periodic motion, as  in the case of a uniform acceleration, Larmor's formula does lead to wrong conclusions. 
Once this fact is realized, much of the doubt or confusion in this long-drawn-out  controversy disappears.
\section{Electromagnetic fields in terms of the ``real-time'' motion of the charge}
It is possible to solve the expression for electromagnetic fields of a uniform accelerated charge, not necessarily in terms of motion of the charge at retarded time as in Eqs.~(\ref{eq:11aaa}), instead wholly in terms of the ``real-time'' motion of the charge \cite{32}. Now, without any loss of generality, we can choose the origin of the coordinate system so that $\alpha=c^{2}/g$, then the position and velocity of the charge at a time $t$ are given by $z_{c}=(\alpha^{2}+c^{2}t^{2})^{1/2}$ and $v=c^{2}t/z_{c}$. Due to the cylindrical symmetry of the system, it is convenient to employ  cylindrical coordinates ($\rho,\phi,z$).
The electromagnetic fields at time $t$ can then be written as \cite{5}
\begin{eqnarray}
\label{eq:32a1}
E_{\rho}&=&8e\alpha^{2}\rho z/\xi^{3}\nonumber\\
\nonumber
E_{z}&=&-4e\alpha^{2}(z_{c}^{2}-z^{2}+\rho^{2})/\xi^{3}\\
B_{\phi}&=&8e\alpha^{2}\rho ct/\xi^{3}\;,
\end{eqnarray}
where $\xi=[(z_{c}^{2}-z^{2}-\rho^{2})^2+4\alpha^{2}\rho^{2}]^{1/2}$.  The remaining  field components are zero. 
Our discussion pertains to the region $z+ct>0$ 
because fields only within this region could have any causal relation with 
the retarded positions of the charge \cite{5}.

The charge happens to be at the same location  at times $t$ and $-t$, i.e., $z_c(t)=z_c(-t)$. Then from Eq.~(\ref{eq:32a1}) it can be seen that the electric field $\bf E$ (with components $E_{\rho},E_{z}$) is an even function of time, i.e., at any given location ($\rho,\phi ,z$), ${\bf E}(t)={\bf E}(-t)$. On the other hand, the magnetic field $\bf B$ (with a component $B_{\phi}$) is an odd function of time, i.e., ${\bf B}(t)=-{\bf B}(-t)$, with ${\bf B}=0$ at $t=0$. Thus there is no Poynting vector, $(\propto {\bf E}\times {\bf B})$, seen anywhere at $t=0$. Further, at any given location, the Poynting vector at time $t$ 
is equal and opposite to its value at $-t$. 

Now at $t=0$, any radiation emitted in past at any time (say, $t=-t_1<0$) should be visible as a Poynting flux passing through a spherical surface of radius $r_1=ct_1$ around the corresponding retarded position of the charge. But the fact that the  Poynting vector is seen {\em nowhere} at $t=0$, implies absence of any radiation emanating from charge at all $t=-t_1<0$. Moreover, corresponding to any event on the charge trajectory even at $t>0$, we can always find an inertial frame which is the instantaneous rest frame of the charge, and in that frame we can thus conclude that no radiation has taken place from the charge at any past event which is in conformity with the assertion that no radiation {\em ever} takes place from a uniformly accelerated charge.
Incidentally, Pauli \cite{33}, exploiting Born's solutions \cite{32}, drew attention to the fact that in the instantaneous rest-frame of a uniformly accelerated charge, ${\bf B}=0$ throughout, and from that he inferred that there might be no
radiation for such a motion.

It has been said in literature that the radiation emitted by a uniformly accelerated (or decelerated) charge goes into regions of space-time inaccessible to a co-accelerating observer \cite{45}. For instance, there are discontinuous $\delta$-fields present 
in the $z = 0$ plane at time $t = 0$, and it is the conjecture that all the radiation emitted by the charge during its uniform acceleration until $t=0$ has gone into these $\delta$-fields \cite{bo80}. However, these $\delta$-fields could have no causal relation with the charge during this period, instead the $\delta$-fields have causal relation with the charge  at time $t= -\infty$ and actually represent the original velocity fields of the charge prior to the onset of acceleration at that time \cite{18}. The energy in the $\delta$-fields could, at most, be
representing the energy loss by the charge due to a {\em rate of change of acceleration} (Eq.~(\ref{eq:6.1})), when the acceleration rose from initial zero value to attain a final constant value, ${\bf g}$, at $t = -\infty$.
It has been explicitly  demonstrated \cite{18} that there is no Poynting flow across the $z=0$ plane at $t>0$ and that the field energy in regions of space-time inaccessible to a co-accelerating observer actually appears at the cost of a steady reduction of energy in $\delta$-fields.

Now, let $\Sigma$ be a fixed finite spherical surface surrounding the charge at $t_1$. The same surface $\Sigma$ surrounds the charge at $-t_1$ as well, because  $z_c(t_1)=z_c(-t_1)$.
The Poynting vector at any point on the surface $\Sigma$ at time $t_1$  
is exactly equal but opposite to its value at time $-t_1$. Therefore Poynting flux through $\Sigma$
\begin{equation}
\label{eq:32a2}
{\cal S}=\int_{\Sigma}{{\rm d}\Sigma}\: ({\bf n} \cdot {\bf S}) 
\end{equation}
at time $t_1$ is equal and opposite to that at $-t_1$. Thus while there may be an {\em outflow} of Poynting flux through surface $\Sigma$ at time $t_1$, but there is an equal 
{\em inflow} of Poynting flux through surface $\Sigma$ at time $-t_1$. This assertion is true for any spherical surface of any radius around the charge, and is therefore not consistent with there being always an outflow of radiation from a surface  surrounding an accelerated (or for that matter even a decelerated) charge, as given by standard radiation formulas.


The electromagnetic field energy in a volume $\cal V$ is given by the volume integral
\begin{equation}
\label{eq:32a3}
\frac{1}{8\pi}\int_{\cal V} {{\rm d}V} \;({E^{2}+B^{2}})\:.
\end{equation}
The field energy density, $({E^{2}+B^{2}})/{8\pi}$, being equal at times $t_1$ and $-t_1$, its volume integral over {\em any chosen} $\cal V$, even in some far-off zone, is also equal at times $t_1$ and $-t_1$. 
Actually, from detailed calculations it has been shown \cite{18} that the total 
electromagnetic field energy in the case of a uniformly accelerated charge, 
including the contribution of the acceleration fields as well, 
at any instant is very much the same as that of a charge moving uniformly
with a velocity equal to the instantaneous ``present'' 
velocity of the accelerated charge. Thus as the charge velocity increases (during the acceleration phase), its net field energy would increase accordingly and the outflow of Poynting flux represents the increasing field energy. On the other hand, as the charge slows down (during the deceleration phase), its net field energy accordingly decreases and the inflow of Poynting flux represents the decreasing field energy in the space throughout.

This deduction is reinforced by the fact that the field energy of the charge is the same at $t_1$ and $-t_1$ since the charge at those two instants is moving with equal speeds (even if in opposite directions). However, there is no trace of any additional energy in electromagnetic fields corresponding to the energy radiated in the intervening period. 

One can also compute the electromagnetic field momentum contained within a volume $\cal V$  from
\begin{equation}
\label{eq:1f1b}
\frac{1}{4\pi c}\int_{\cal V}{{\rm d}V}\:({\bf E}\times{\bf B}).
\end{equation}
Since $\bf B=0$ at $t=0$ (Eq.~(\ref{eq:32a1})), there is no momentum in the electromagnetic fields anywhere, in the instantaneous rest frame.

Further, for  $t\ne0$, from Eq.~(\ref{eq:1f1b}) in conjunction with Eq.~(\ref{eq:32a1}), the electromagnetic field momentum within {\em any chosen} $\cal V$, even in some far-off zone, is not only equal but in opposite directions at times $t_1$ and $-t_1$. Now, this fits the expectation that since the charge occupies the same location but has equal and opposite velocities at $t_1$ and 
$-t_1$, any given volume element should contribute to the self-fields equal and opposite field momentum, proportional to the ``present'' velocity, at $t_1$ or  $-t_1$. From the cylindrical symmetry of electromagnetic fields (Eq.~(\ref{eq:32a1})), it is easily seen that the net electromagnetic field momentum, integrated over all space, is directly proportional to the instantaneous velocity $\bf v$ of the charge. However, there is 
no trace of any additional momentum being carried away by radiation fields which would even otherwise have different angular distributions, due to Doppler beaming along opposite directions of velocities, at $t_1$ and $-t_1$. 
\section{Conclusions}
We showed that contrary to the predictions of Larmor's/Li\'{e}nard's radiation formula, no observer anywhere would detect any radiation from a uniformly accelerated charge, in agreement with the absence of electromagnetic radiation from such an  accelerated charge. This conclusion is also in accordance with the strong principle of equivalence.
{}

\begin{thebibliography}{}
\bibitem{la97} J. Larmor, {Phil. Mag.} {44} (1897) 503
\bibitem{1} J. D. Jackson, {Classical electrodynamics}, 2nd ed. (Wiley, New York, 1975)
\bibitem{2} W. K. H. Panofsky and M. Phillips, {Classical electricity and magnetism}, 2nd ed. (Addison-Wesley, Massachusetts, 1962)
\bibitem{25} D. J. Griffiths, {Introduction to electrodynamics}, 3rd ed. (Prentice, New Jersey, 1999)
\bibitem{68d} A. K. Singal, {J. Phys. Commun.} {2} (2018) 031002
\bibitem{4} C. W. Misner, K. S. Thorne and J. A. Wheeler,  {Gravitation} (Freeman: San Fransisco, 1973)
\bibitem{28} A. K. Singal, {Am. J. Phys.} {79} (2011) 1036-1041
\bibitem{17} A. K. Singal, {Gen. Rel. Grav.} {27} (1995) 953-967
\bibitem{18} A. K. Singal, {Gen. Rel. Grav.} {29) (1997) 1371-1390
\bibitem{68a} A. K. Singal, {Eur. J. Phys.} {37} (2016) 045210
\bibitem{abr04} Abraham M., { Ann. Phys.} {14} (1904) 236
\bibitem{abr05} M. Abraham, {Theorie der elektrizitat, Vol II: Elektromagnetische theorie der strahlung} (Teubner, Leipzig, 1905)
\bibitem{lor04} H. A. Lorentz, {Encykl. Mathe. Wiss.} {2} (1904) 145-280
\bibitem{16} H. A. Lorentz, {The theory of electron} (Teubner, Leipzig, 1909); 2nd ed. (Dover, New York, 1952)
\bibitem{68b} A. K. Singal, {Am. J. Phys.} {85} (2017) 202-206
\bibitem{32} M. Born, {Ann. Physik} {30} (1909) 1-56
\bibitem{5} T. Fulton and F. Rohrlich, {Ann. Phys.} {9} (1960) 499-517
\bibitem{33} W. Pauli  {Relativit{\"a}tstheorie} in Encyklopadie der Matematischen Wissenschaften, V {19} (Teubner, Leipzig, 1921)
Translated as {Theory of relativity} (Pergamon, London, 1958)
\bibitem{45} C. de Almeida and A. Saa, {Am. J. Phys.} {74} (2006) 154-158
\bibitem{bo80} D. G. Boulware, {Ann. Phys.} {124} (1980) 169-188}
\end{thebibliography}
\end{document}